\documentclass[5p]{elsarticle}

\pdfoutput=1

\usepackage[utf8]{inputenc}
\usepackage{units}
\usepackage{amsmath}
\usepackage{url}
\usepackage{graphicx}
\usepackage{subfigure}
\usepackage{booktabs}
\usepackage{floatrow}
\usepackage{microtype}
\usepackage{hyperref}
\usepackage{csquotes}

\newcommand{\hess}{H.E.S.S.}
\newcommand{\hessI}{\hess{}~I}
\newcommand{\hessII}{\hess{}~II}
\newcommand{\phaseI}{Phase~I}
\newcommand{\phaseII}{Phase~II}
\newcommand{\hessphaseI}{\hess{}\ \phaseI{}}
\newcommand{\hessphaseII}{\hess{}\ \phaseII{}}
 
\begin{document}
\epstopdfsetup{suffix=}

\title{ The \hess{} central data acquisition system}

\author[desy,potsdam]{A. Balzer}
\ead{arnim.balzer@desy.de}
\author[potsdam]{M. Füßling}
\author[hu]{M. Gajdus}
\author[ecap]{D. Göring}
\author[potsdam]{A. Lopatin}
\ead{anton.lopatin@fau.de}
\author[llr]{M. de Naurois}
\author[cern]{S. Schlenker}
\author[hu]{U. Schwanke}
\author[potsdam,desy]{\mbox{C. Stegmann}}

\address[desy]{DESY, D-15738 Zeuthen, Germany}
\address[potsdam]{Institut f\"ur Physik und Astronomie, Universit\"at Potsdam,  Karl-Liebknecht-Strasse 24/25, D-14476 Potsdam, Germany}
\address[hu]{Institut f\"ur Physik, Humboldt-Universit\"at zu Berlin, Newtonstr. 15, D-12489 Berlin, Germany}
\address[ecap]{Physikalisches Institut, Universit\"at Erlangen-N\"urnberg, Erwin-Rommel-Str. 1, D-91058 Erlangen, Germany}
\address[llr]{Laboratoire Leprince-Ringuet, Ecole Polytechnique, CNRS/IN2P3, F-91128 Palaiseau, France}
\address[cern]{CERN, CH-1211 Geneva 23, Switzerland}

\begin{abstract}
The High Energy Stereoscopic System (\hess{}) is a system of Imaging Atmospheric Cherenkov Telescopes (IACTs) located in the Khomas Highland in Namibia. It measures cosmic gamma rays of very high energies \mbox{(VHE; $>\unit[100]{GeV}$)} using the Earth's atmosphere as a calorimeter. The \hess{} Array entered \phaseII{} in September 2012 with the inauguration of a fifth telescope that is larger and more complex than the other four. This paper will give an overview of the current \hess{} central data acquisition (DAQ) system with particular emphasis on the upgrades made to integrate the fifth telescope into the array. At first, the various requirements for the central DAQ are discussed then the general design principles employed to fulfil these requirements are described. Finally, the performance, stability and reliability of the \hess{} central DAQ are presented. One of the major accomplishments is that less than $\unit[0.8]{\%}$ of observation time has been lost due to central DAQ problems since 2009.
\end{abstract}


\maketitle
\tableofcontents


\section*{Introduction}
\noindent
The High Energy Stereoscopic System (\hess{}\footnote{\url{http://www.mpi-hd.mpg.de/hfm/HESS/}}) is an array of Imaging Air Cherenkov Telescopes (IACTs) located in the Khomas Highland of Namibia. It is dedicated to the observation of very-high-energy (VHE) gamma rays \citep{crab}. IACTs detect Cherenkov light emitted by charged particles in cosmic-ray-initiated air-showers. The measured light distributions are then used to reconstruct the properties of the primary particle. More details about the IACT technique, and the state of gamma-ray astronomy in the \unit{TeV} regime can be found in \citep{TeVastro}.

Currently, the \hess{} experiment has entered so called “\phaseII{}” with the inauguration of a fifth telescope \citep{hess2Vincent,hess2Punch} on September $28^\text{th}$, 2012. This upgrade of the \hess{} Array will lower the energy threshold of the experiment from about $\unit[100]{GeV}$ \cite{crab} to about $\unit[30]{GeV}$ \cite{HESSIIBecherini,HESSIIHoller} and improve the overall sensitivity by a factor \mbox{of $\sim2$} in the central energy range.

In the course of the upgrade, the central data acquisition (DAQ) system of the array \citep{daq} was also revised. The on-site computing cluster (next to the telescopes) was replaced by modern off-the-shelf hardware while the DAQ software was ported to this new environment and adjusted to meet the requirements of \hessphaseII{}.

This paper gives an overview of the central DAQ system and the challenges that were met and overcome during the cluster upgrade. Unlike the DAQ systems of VERITAS \citep{VERITAS} and MAGIC \citep{MAGIC}, the \hess{} DAQ system is also responsible for array control and monitoring. We will therefore address these topics and also mention error propagation within the system, as well as the user interaction with the DAQ. Furthermore, the reliability and performance of the central DAQ are discussed. Finally, a summary of the important lessons learned during the process is given, which might be helpful to upcoming ground-based gamma-ray telescope arrays like the Cherenkov Telescope Array (CTA) \citep{CTAdesignReport}.


\section{Requirements}
The requirements of a DAQ system for an experiment like \hess{} can be divided into three groups. First, there are design goals that need to be met in order to achieve optimal scientific output. Moreover, there are technical requirements that need to be fulfilled by the DAQ. The last but not least group of requirements is about ensuring that the system is as user-friendly as possible.

\subsection{Science requirements}
The expected flux of VHE gamma rays observable by ground-based IACTs like \hess{} is very low compared to the expected background rate. The IACT arrays use stereoscopic observation of air showers to help reduce the high background rate of hadron-induced air showers as well as to improve the direction reconstruction of photon-induced air showers \citep{crab}. Therefore, the dead time of the whole array must not be increased by the DAQ. As a result, the number of telescopes that detect an air shower and are able to record the event is not to be reduced by the DAQ. Moreover, variations in the trigger rate such as fluctuations induced by night sky background or by transient phenomena (i.e.~bursts) should not cause dead time due to DAQ related processing back-pressure.

The maximum hardware event readout rate is about $\unit[900]{Hz}$ with an event size of $\unit[4.5]{kB}$ for each of the \hessI{} telescopes (CT1-4) and about $\unit[3.4]{kHz}$ with an event size of $\unit[10]{kB}$ for the fifth telescope (CT5) \citep{CT5camera}. This leads to maximum data rates of approx.\ $\unitfrac[50]{MB}{s}$ for the primary scientific data during routine operation. To account for other hardware devices on-site as well as some additional capacity for tests, the DAQ system is required to process at least $\unitfrac[80]{MB}{s}$, thereby having sufficient throughput for network connections and storage facilities. The same data format, as decided by the \hess{} Collaboration, is used for storage on-site as for high level tasks like event reconstruction and data analysis. This approach allows high level analysis algorithms and visualisation tools to be used on a DAQ level. Therefore, the raw data that are sent from the Cherenkov detectors have to be converted into a common data format at a very early stage, which requires additional computing power. Moreover, the common data format allows data that are identical to those recorded by the DAQ to be easily simulated. 

On top of the mere acquisition of data, their integrity and quality needs to be verified in real time. Furthermore, the data should be checked and analysed by a preliminary analysis during observation. Due to the common data format it is possible to use the standard offline analysis software for this purpose. The array needs to be able to operate in different observation modes with different sets of telescopes. Several independent operation modes with disjoint sets of telescopes, so called \emph{SubArrays}, should be possible simultaneously, for example, \emph{observation} runs as well as \emph{calibration} and \emph{maintenance} runs.

The DAQ system needs to be able to respond to target of opportunity (ToO) alerts, like alerts from the Gamma-Ray Burst Coordinates Network\footnote{The Gamma-Ray Burst Coordinates Network (GCN) distributes information about the location of a Gamma-Ray Burst detected by various spacecraft \url{http://gcn.gsfc. nasa.gov/}}, in real time. In order to allow for prompt follow-up observations, the DAQ should respond to such alerts as fast as possible (including a slewing of the telescopes to a new observation target) and should thus be highly automated. On top of that, the DAQ must be able to handle continuous data streams of other hardware components (for example monitoring data from weather stations) as well as to merge these streams with the data taken during observation.

The \hess{} Array only takes data during the moonless part of a night when the weather conditions are good, i.e.~no clouds are above the detector. To maximise the physics output of these time periods the scheduling of the observation targets should be automated and configurable remotely (i.e. from Europe where most host institutes are).

\subsection{Technical requirements}
To facilitate the communication between the different devices, subsystems and computers, a common network infrastructure has to be established. This network should be based on robust, proven and commonly available technologies. This, in general, also holds for the hardware of the on-site computing cluster. So, whenever possible, standard off-the-shelf components should be used for equipment like computing nodes, storage servers and desktop PCs. This reduces the cost and simplifies the replacement of broken components.

To account for future developments and upgrades, the DAQ design should be scalable in terms of processing power, data throughput and storage capacity. Furthermore, it should be possible to incorporate new hardware devices easily, i.e.~with no change to the general data storage and data transportation structures. All vital DAQ system components should be redundant to allow exchanging broken hardware components quickly to minimise loss of observation time due to DAQ problems. In case of a power cut the DAQ system should remain operational and prevent data corruption.

Each device in the array must be mapped to at least one controlling software process (\emph{Controller}). More complex hardware devices may correspond to multiple \emph{Controllers}. Furthermore, the configuration of the different pieces of hardware, as well as the DAQ itself, should be as flexible as possible. New hardware should be usable without the need to change major parts of the code (e.g.~data transport and device monitoring). As a side effect of a flexible configuration, the array is more tolerant to missing or malfunctioning hardware.

To prevent damage to the telescopes, the Cherenkov cameras and other critical systems, error handling needs to be redundant and decentralised. Therefore, each subsystem (e.g.~the camera or the drive system) is responsible for its own safety and fatal problems have to be handled directly in firmware. Only after immediate danger is averted, the devices inform the DAQ system of the error condition, which in turn takes appropriate actions, such as propagating the error to other subsystems or bringing the corresponding \emph{SubArray} to a safe state. In order to avoid hazardous conditions for the hardware, slow control information from all the subsystems (e.g.~device temperatures, positions or voltages) has to be monitored by the DAQ.

The \hess{} Array is situated in a remote location where no cheap, reliable and fast internet connection is available. Therefore, the data cannot be streamed to the different institutes of the \hess{} Collaboration. Instead, the data have to be shipped to the institutes in regular intervals via magnetic tapes. The DAQ provides the resources to store the data on-site for up to three months until the shipped data has been verified in Europe.

In general, the system should be as self-sufficient as possible. There must be no dependencies of DAQ components on remote network connections whatsoever, meaning data taking must not be interrupted by a failing internet connection. Furthermore, all relevant documentation and manuals need to be available on-site.

\subsection{User requirements}
The \hess{} DAQ should facilitate integration of new hardware and allow a quick response to hardware malfunctions without interfering with the rest of the array. The software should, therefore, be modular and support different hardware configurations.

The \hess{} Array is operated by collaboration members (PhD students,\ldots) during "Shifts" of one moon period. As a result, the array is operated by monthly changing non-expert personnel on-site (Shifters) with remote support by a team of subsystem experts working at the respective member institutes as well as a local shift expert responsible for the training of the Shifters at the beginning of each Shift. As a security precaution the access of Shifters to critical parts of the array has to be limited (for example configuration databases or low-level telescope movement), while subsystem experts must be able to perform remote maintenance. Moreover, the DAQ software should use stable releases that may not be altered by Shifters for observations, but should still allow subsystem experts on-site to develop and debug their software modules.

As subsystem experts are working at their home institutes most of the time and not on the \hess{} site, a remote access to the various subsystems of the array must be available. Due to limitations in internet connectivity, these remote connections need to be simple. To further ease the maintenance of the array, a dedicated logging system with detailed information for Shifters and subsystem experts has to be part of the DAQ software. For testing purposes, and in case of malfunctioning hardware, the ability to manually override automated procedures must be given at all times.

Due to the non-expert personnel operating the array during a shift, manuals and documentation have to be available on-site at all times. This also includes detailed instructions about safety regulations and error recovery. The DAQ should also be operable in a semi-automatic way by a single Shifter (for security reasons at least two Shifters have to be present at all time). Moreover, the Shifters must be able to get fast feedback about the current status of the array during data taking and especially in the case of system errors. Easy to understand error reports and clearly described error recovery procedures are key to high-efficiency data taking. By design, the Shifters have the final say in deciding whether data taking continues or not after an error has occurred. This also holds true in \hess{} for weather and atmospheric conditions: there is no automated response from the DAQ and the Shifters' decision is necessary.


\section{Central DAQ hardware}

\begin{figure*}[tb]
\floatbox[{\capbeside\thisfloatsetup{capbesideposition={right,top},capbesidewidth=4cm}}]{figure}[\FBwidth]
{\caption[]{Scheme of the network layout on the \hess{} site in Namibia. In the upper part, the server room with the five storage servers as well as the ten worker nodes is depicted. Furthermore, the Central Trigger as well as the main switches are shown. The Cherenkov Telescopes (CT) are labeled from one to five, with the last one being the \hessII{} telescope. Black lines with arrows indicate gigabit data network ethernet connections while the green lines with circles represent a physically separated gigabit ethernet network for the mounting of the NFS and GlusterFS servers. The purple lines with the diamond shaped ends represent the direct optical fibre connections between the Cherenkov camera trigger sytems and the Central Trigger.}\label{Overview}}
{\includegraphics[width=350px]{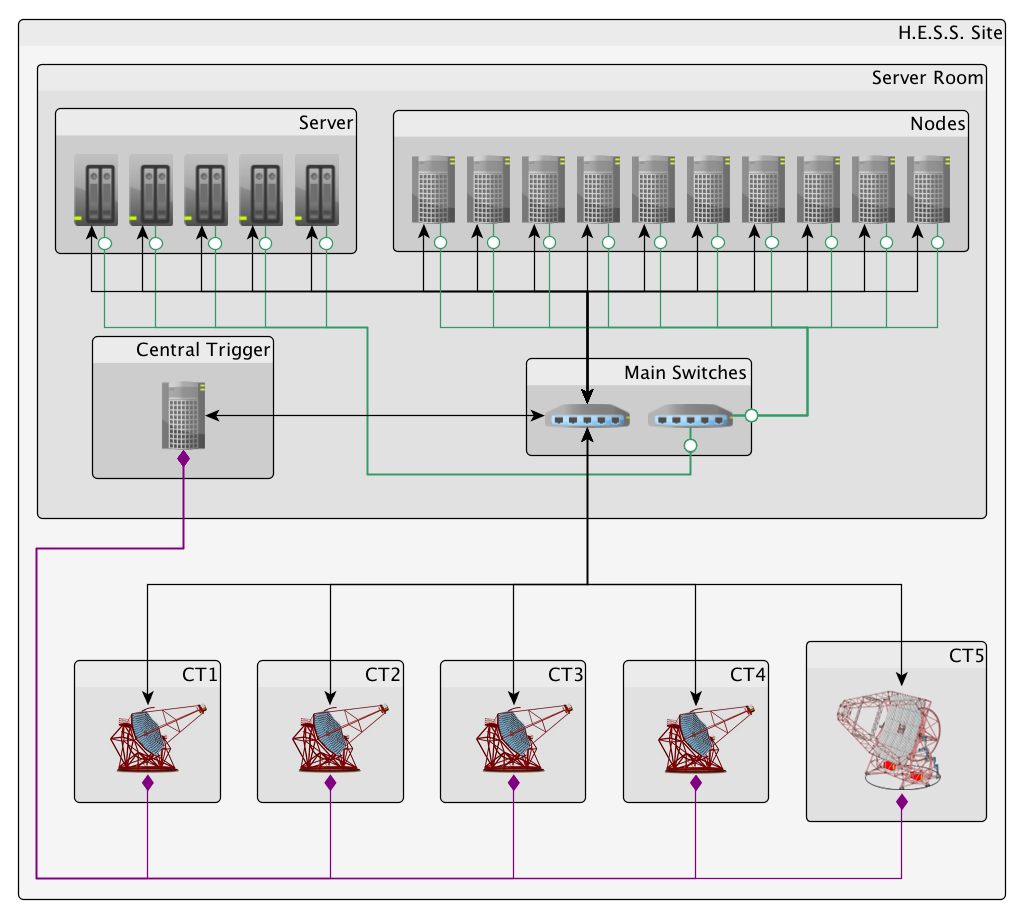}}
\end{figure*}

The primary data flow starts in the Cherenkov cameras of the five telescopes. If an air shower event is seen by a telescope, i.e.~the camera trigger decided to launch the read out of an event, the camera sends a trigger signal via optical fiber to the Central Trigger and begins to digitise the recorded image. The Central Trigger checks for coincidences in at least two telescopes, if no coincidence is found it sends a \enquote{fast-clear} signal to the telescope, which will drop the event. In case at least two telescopes trigger within $\approx\unit[80]{ns}$, the camera image is sent to the DAQ. A more detailed discussion of the \hessII{} Central Trigger and camera readout can be found in \citep{CT5camera,hesstrigger}. All the data recorded by the different cameras that belong to one event are sent via network to one node (see Figure~\ref{Overview}) in the \hess{} DAQ cluster. Furthermore, the data from the Central Trigger are sent to the same node, including a unique event number and GPS time stamp for each event. The received raw data are buffered in memory at the receiving node, converted to a common data format by that node and stored on file servers.

\subsection{Network implementation}

The communication between the different components of the DAQ is done via a Gbit Ethernet network only. If a piece of equipment is not able to use Ethernet, a gateway server is used, e.g.~a serial COM-Server. Several Gbit switches connect all hardware with each other on the \hess{} site via two 48 Port switches as the central communication hub inside the server room. Long distances are covered using Gbit media converters and optical fibers. The connection between the camera trigger systems and the Central Trigger is a direct optical fibre connection and not part of the normal data network. This is done to achieve minimal signal latency to reduce the over all dead time of the array. The average mean dead time during normal observation is $\approx \unit[8]{\%}$ for the whole array ($\approx \unit[7]{\%}$ for CT5 and $\approx \unit[9.3]{\%}$ on average for the others). For the dead time of the camera electronics see \citep{CT5camera}. Another exception are the distributed file systems (see Section \ref{storage}) used within the \hess{} DAQ which are accessed via an additional physically separate Gbit Ethernet network so that they can't interfere with data taking.

\subsection{Computing \& storage servers}\label{storage}

The cluster of the \hessII{} Array consists of ten worker nodes and five storage servers. It is located in a climate controlled server room inside the control building next to the telescopes. Each storage server is equipped with a $\unit[12]{TB}$ RAID6 \citep{wiki:raid6} with an additional hot spare disk using an XFS \citep{XFS} file system. This ensures redundancy as well as enough IO bandwidth and disk space for data taking. The NFS \citep{NFS} and GlusterFS \citep{GlusterFS} protocols are used for distributed file access through the worker nodes. Due to the low-bandwidth internet connection on-site, data have to be transported using magnetic tapes at the end of each Shift, currently LTO-4 tapes \citep{wiki:lto} are used. The storage capacity on-site is sufficient to hold all data until it has been received and verified in Europe which does not take longer than 3 months. Every month \hessphaseI{} takes $\unit[420]{GB}$ of data and for \hessphaseII{} this value is around $\unit[11]{TB}$, while, the total available disk space on-site is about $\unit[60]{TB}$.

Different hardware subsystems require special purpose machines, like e.g.~custom boot servers for the camera and Central Trigger board electronics. These custom machines have been developed together with their respective hardware components in the lab. On-site they have been turned into virtual machines which are hosted by one of the servers. These virtual machines are fully customized to their respective task, e.g.~using $\unit[32]{bit}$ CPU architecture. Furthermore, it is possible to backup the machines and migrate them to another physical host in case of a hardware failure.

\subsection{Cluster monitoring \& maintainability}
The maintenance of the \hess{} central DAQ computing cluster and the DAQ software itself is crucial for the operation of the experiment. To facilitate the maintenance and to reduce the complexity of the overall system, the cluster was designed to be as homogeneous as possible. This means that the different computing nodes and storage servers are based on the same hardware and run the same operating system allowing the same spare parts and same software to be used for all machines. To cope with power interruptions, which are quite common at the \hess{} site (several times a week), a diesel generator in combination with online/double-conversion uninterruptible power supplies (UPSs) \citep{wiki:ups} are used to ensure a continuous power supply to the cluster. The diesel generators provide enough power for the telescopes and all electronic equipment on-site. However, since they need several seconds to kick in, the UPS are needed to ensure a constant power output to the computing cluster. If necessary, the DAQ cluster can be sustained by the UPS for up to 25 min. The status of the cluster is monitored constantly with several open source tools like \emph{Ganglia} \citep{ganglia} and \emph{Collectd} \citep{collectd}. On top of that, email notifications as well as \emph{SNMP} traps are used to get error notifications from the different hardware components.

\subsection{Software maintenance \& backups}

A uniform administration of all machines in the cluster is possible, using the same software packages. The \emph{System Imager} tool \citep{systemimager} is used to create daily backups of the current operating system, including all relevant software to run the \hess{} central DAQ. If the operation system is flawed after a software change, the System Imager allows the system to be rolled back to a previous, stable image with a single command. Dedicated backups of important files are performed on a daily basis. With increasing age of the backups, the frequency is reduced from daily, to weekly, to monthly snapshots. The databases (see section \ref{software:database}), that are being used throughout the DAQ, are also backed up on a daily basis.


\section{DAQ software concept and implementation}

\subsection{Data format}\label{sec:dataformat}
The transport and storage of objects needs a serialisation mechanism which is provided by the ROOT Data Analysis Framework \citep{ROOT} on which the \hess{} raw data format is based.

To ensure a common interface and easy access to all the different raw data formats used in \hess{}, a common raw data base class is implemented in the software. This base class also takes care of correct time stamps for the different events that are recorded and saved to disk. This enforces a correct order in the loading and processing of events. Both the ROOT-based \hess{} data format and the ROOT graphics and histogram classes are used online and off-line allowing a seamless integration of the high-level data analysis code into the DAQ software, see Section \ref{RealTimePipeline}.

\begin{figure}[!tb]
\includegraphics[width=\columnwidth]{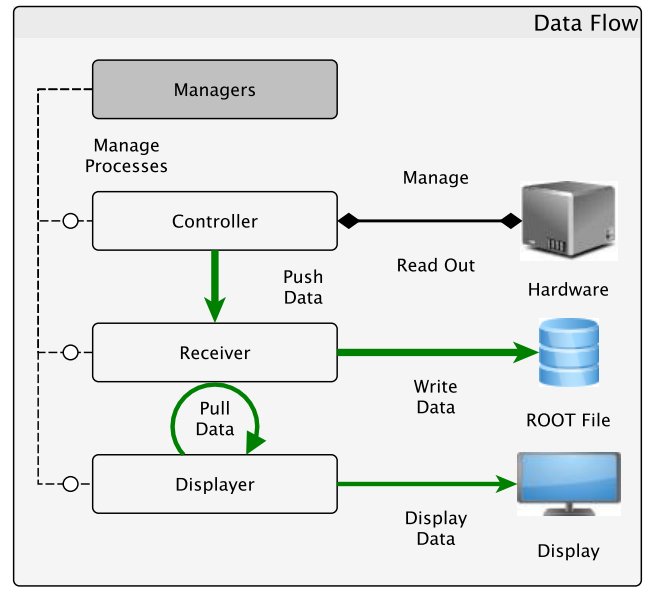}
\caption[]{Data flow within the \hess{} DAQ. Each piece of hardware is read out and managed by at least one DAQ \emph{Controller} process. The data which is obtained from the hardware is sent using the \emph{Push} mode to the corresponding receiver process. In \emph{Push} mode data integrity  is guaranteed. The received data is then stored in ROOT files using the common \hess{} data format. For fast feedback to the Shift Crew, the data processed by a receiver can be pulled by a \emph{Displayer} process and displayed on one of the available screens in the control room. The data can be sent at an arbitrary rate. Error handling and process synchronisation is taken care of by the \emph{Manager} process.}
\label{DataFlow}
\end{figure}

\subsection{Data transport}\label{Corba}

The communication between the different processes in the DAQ is built on the CORBA \citep{CORBA} distributed, object-oriented inter-process communication standard. A client process can call remote procedures on a server object using CORBA messages. The result of the function call (with possible read, read-write and write parameters) is propagated back to the client after the execution of the code on the server side. Due to the combined usage of CORBA and ROOT, several multi-threading issues in the ROOT Framework were discovered. To achieve the required thread safety, several patches have been contributed to ROOT, and are now part of all ROOT releases \citep{RootChangeLog, RootPresentation}.

CORBA allows the use of different programming languages and different operating systems on the server and the client side. The \hess{} DAQ uses the free (LGPL) omniORB \citep{omniORB} implementation, which comes with C++ and Python support. A central directory naming service provides object registration (using a hierarchical naming tree with processes being grouped by CORBA contexts) and allows any process to easily obtain connections to other remote processes.

The \hess{} DAQ implements two different data transport mechanisms: pushing and pulling of data, see Figure~\ref{DataFlow}. In \emph{Push} mode the data are sent from one process to another, ensuring the data reception and integrity. The \emph{Push} mode is used to store all measured data, including the scientific data from the Cherenkov cameras. In \emph{Pull} mode, a client polls the server process for new data in periodic intervals. Like this, data can be dropped or requested at a lower rate, but the data integrity is still ensured. This transport mechanism is used by displaying processes where a sub-sample of the data is sufficient.

To store data on disks, multiple instances of a generic receiver process are used throughout the DAQ. These receiver processes are implemented in a generic way which allow the introduction of new data without the need to change existing code. The generic receiver makes use of the common base class of the \hess{} raw data formats and can save any data using the ROOT serialisation mechanism. Moreover, basic data quality checks can be performed while the data is being received, e.g.~monitoring if the time that passes between two events or the size of the event is in a certain range and has a certain mean value.

\subsection{Node switching}

To be able to deal with high data rates so called \emph{Node Switching} is used in the \hess{} DAQ, this is a round-robin load balancing scheme \citep{wiki:roundrobin}. All the telescopes in a given run send their data to one node for a given amount of time (usually $\unit[4]{s}$). After that, the Central Trigger announces the next node that is free to receive data from all the cameras. In case of insufficient computing power the number of nodes can be adjusted manually. Using this scheme the DAQ can receive and process data at higher rates than required by the \hess{} telescope array.

A specialisation of the generic receiver mentioned before is the \emph{CameraReader}, which is the process responsible for receiving, buffering and processing the Cherenkov camera events. The received events are unpacked, converted into the \hess{} raw data format, joined with the Central Trigger information and finally written to the storage servers. For very high data rates, the storage servers may not be able to store the events fast enough. In that case the nodes can be configured to use their local disks for the duration of the observation after which the data are merged and copied to the storage servers.

\subsection{Controllers \& state machine}

All hardware used by the DAQ is represented by a \emph{Controller}, which is a software process running on a computing node on the DAQ cluster, acting as a uniform interface between the given hardware component and the DAQ itself. Simple hardware can be represented by a single \emph{Controller} whereas more complex hardware like the Cherenkov cameras use multiple \emph{Controllers}, e.g.~\emph{Camera HV Controller}, \emph{Camera Trigger Controller}, \emph{Camera Lid Controller}, $\dots$ A state machine maps the state of the hardware to the state of the corresponding \emph{Controllers}. The mapping of \emph{Controller} states to hardware states for some processes is shown in Table~\ref{tab:States2hardware}. If a \emph{Controller} is in the \emph{Safe} state, the corresponding hardware is either turned off or in a state of minimal activity. In preparation for data taking and in between runs all \emph{Controllers} are in the \emph{Ready} state, i.e.~the hardware is turned on and slow control information is being recorded. As an intermediate step shortly before data taking, the \emph{Configured} state indicates that the hardware has received all the necessary configuration parameters from its \emph{Controller}. Finally, if a \emph{Controller} is \emph{Running} the corresponding hardware is read out and the data are processed by the DAQ and stored on disk.

\begin{table*}[!htb]
  \centering
  \caption{Mapping of state machine states to real hardware states for some important software processes in the \hess{} DAQ.}
  \begin{tabular}{l c c c c }
    \toprule    
    & \multicolumn{4}{c}{Controller state} \\ \cmidrule(r){2-5}
    Process & Safe & Ready & Configured & Running \\
    \midrule
    Node Receiver & Idle & Buffers allocated & Camera Configuration Read & Receiving \& Saving Data \\
    Camera & Power Off & Power On, Cooling On & Drawers configured & Reading Data\\
    Camera HV & Off, $\unit[0]{V}$ & On, $\unit[800]{V}$ & On, Full Voltage & On, Full Voltage \\
    Camera Lid & Closed &  Open & Open & Open \\
    Camera Trigger &  Off & Off & Input Configured & Trigger Active \\
    Tracking & Parked In & Parked Out & On Target & On Target \\
	\bottomrule
  \end{tabular}
  \label{tab:States2hardware}
\end{table*}

\begin{figure*}[!htb]
\includegraphics[width=\textwidth]{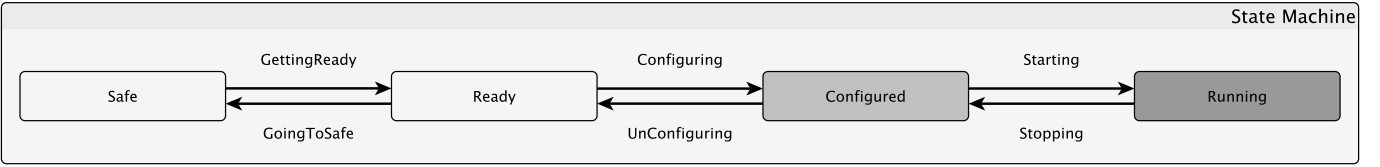}
\caption[]{Visualisation of the state machine used in the \hess{} DAQ. The boxes represent the states a \emph{Controller} (and its respective device) can be in, while the arrows show the available transitions. This state machine is linear by design, which simplifies the synchronisation of multiple \emph{Controllers}. As all the hardware \emph{Controllers} are mapped to the same state machine, it becomes quite easy to determine the state of the whole array or any subset, e.g.~when all processes of telescope CT1 are \emph{Ready}, the whole telescope is \emph{Ready} for observation.}
\label{StateMachine}
\end{figure*}

State transitions are used to change from one state to another. A scheme of the state machine used by the \hess{} DAQ can be seen in Figure~\ref{StateMachine}. If a \emph{Controller} cannot perform its state transition successfully (for example due to hardware problems) it will fall back to its previous state. A detailed description of the error handling is given in Section \ref{ErrorHandling}.

\begin{figure*}[!htb]
\includegraphics[width=\textwidth]{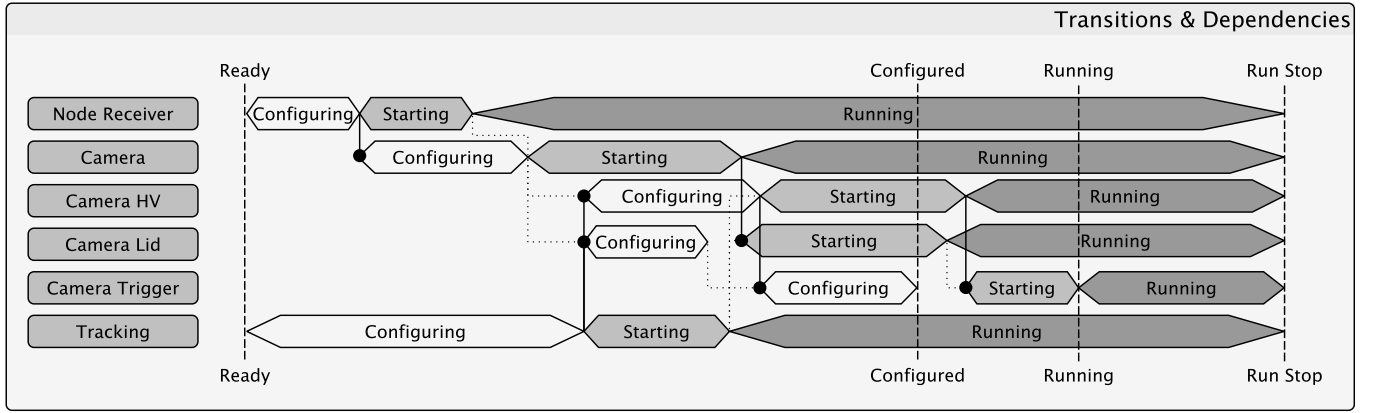}
\caption[]{An example interaction of various important \hess{} DAQ processes during the starting of an observation run. The change of the state of the corresponding \emph{SubArray} is indicated by the dashed lines and ranges from \emph{Ready}, over \emph{Configured} to \emph{Running}. The hexagonal boxes represent state transitions of the corresponding DAQ \emph{Controllers}. The solid and dotted black lines with a filled circle at the end indicate dependencies between processes within the \hess{} DAQ, i.e.~the HV of the PMTs in the camera must not be turned on while the telescope is still moving and the camera trigger should not be activated before the camera HV is turned on. As a result a process only starts with its transition if all of its dependencies have successfully finished their own corresponding transition. The solid lines indicated the slowest dependency of a given process while the dotted lines represent dependencies that already reached the required state. The global state of the \emph{SubArray} is determined by the slowest process, i.e.~once the last process finished its transitions from \emph{Ready} to \emph{Configured} the whole \emph{SubArray} is considered to be \emph{Configured}.}
\label{Transitions}
\end{figure*}

There is only one transition from one state to an adjacent state for a given direction which simplifies the handling of the state machine because loops or unreachable states cannot occur. Furthermore, a flat state machine allows processes to be sent to any given target state that need not be adjacent to the current state, i.e.~from \emph{Safe} to \emph{Running} or vice versa. (The \emph{Controller} will perform all necessary transitions to get from the \emph{Safe} to the \emph{Running} state automatically, i.e.~the order of executed transitions would be \emph{GettingReady}, \emph{Configuring} and \emph{Starting}.) Transitions towards the \emph{Running} state are called “upwards”, and those in the other direction are called “downwards”. To be able to monitor the performance of the DAQ processes, time stamps are written into a MySQL database at the beginning of each transition, once all of the dependencies (see Figure~\ref{Transitions}) of a \emph{Controller} have finished their transition and again at the end of each transition (see Section \ref{tt_tools}).

\subsection{Managers \& dependencies}\label{Manager}
\begin{figure*}[tb]
\floatbox[{\capbeside\thisfloatsetup{capbesideposition={right,top},capbesidewidth=4cm}}]{figure}[\FBwidth]
{\caption[]{An example overview of the process hierarchy within the \hess{} DAQ is shown. The \emph{Run Manager} distributes the scheduled data taking runs over an arbitrary number of \emph{SubArrays}. Each \emph{SubArray} consists of a \emph{SubArray Manager} process and several other sub contexts. These contexts normally are various Node and CT contexts representing the Cherenkov telescopes and the \emph{CameraReaders}. Each \emph{SubArray} can operate independently from the others and Nodes and CT contexts can be assigned in any combination to any \emph{SubArray}. The so-called Slow Control context is responsible for atmospheric monitoring tasks as well as all processes that display data to the Shift Crew and is constantly in a \emph{Running} state.}\label{Hierarchy}}
{\includegraphics[width=350px]{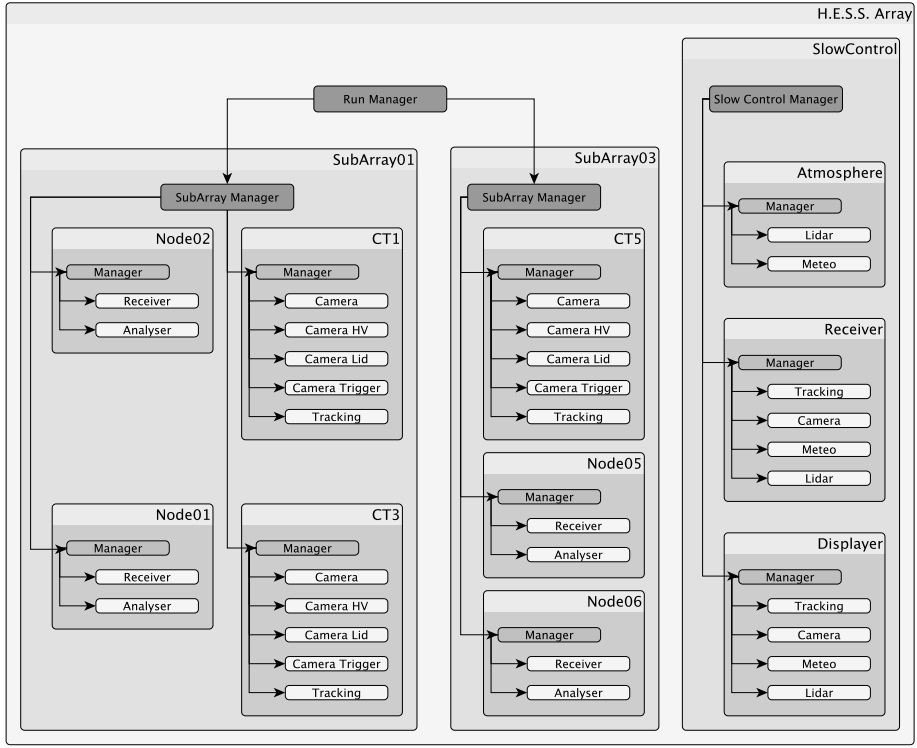}}
\end{figure*}

To synchronise the different processes during data taking \emph{Managers} are used throughout the DAQ. Each \emph{Manager} is also a \emph{Controller}, i.e.~a software process running on a computing node of the DAQ cluster, providing an extended interface to deal with collections of \emph{Controllers}. The \emph{Managers} are responsible for distributing the run configuration to their subordinated processes as well as managing their states. The state of the Manager is determined by the minimum state of the all managed processes, e.g.~if the CT5/Tracking is \emph{Running} (i.e.~CT5 is tracking the observation target) but some camera process in CT5 is still \emph{Starting} (e.g.~the high voltage is turned on) the CT5/Manager's state is still \emph{Configured}. Furthermore, they are used for error handling which is discussed in detail in Section \ref{ErrorHandling}. The DAQ uses a hierarchical structure to control all the processes of a given run; with the \emph{SubArray Manager} on top and \emph{Managers} for every Context (see Section \ref{Corba}) in the order of the hierarchy of the contexts. The hierarchy of \emph{Managers} and \emph{Controllers} in the \hess{} DAQ is shown in Figure~\ref{Hierarchy}.

The \emph{Manager} of a given context gets the list of the processes needed for a given run from the database. There are three distinct types of processes that can be specified within the database. \emph{Disturbing} processes are sent to a safe state at the beginning of the run, e.g.~the camera lid is disturbing for a park-in run. Therefore, it is closed before the telescopes are parked in (i.e.~moving the camera into the camera shelter and closing its roof). \emph{Required} processes are necessary for a given run and a run will be stopped or will fail to start if a required process is not in the correct state. Finally, \emph{Optional} processes are part of a given run but an error or a misoperation of one of these processes does not influence the data taking, e.g.~a crash of the process displaying monitoring data to the Shifters does not stop data taking. Once the problem with an optional process is fixed it is allowed to rejoin data taking.

In certain cases, a \emph{Controller} has to be in a certain state before another \emph{Controller} can begin to perform its transition to its target state. These dependencies are entered into the database and \emph{Controllers} will get a list of all their dependencies for a given run type from their corresponding \emph{Manager}. \emph{Controllers} will wait for their dependent processes to finish their transitions before starting with their own, e.g.~a camera trigger will wait for the camera HV to be turned on before configuring itself. Some example dependencies of processes within the DAQ can be seen in Figure~\ref{Transitions}. Dependencies are applied in reverse order for downwards transitions, this ensures that processes are shut down in the right order.

\subsection{Configuration database}\label{software:database}

In general, the whole configuration of the \hess{} DAQ is contained in MySQL databases, including the processes and machines in use, as well as common environment variables and the configuration of the various hardware components. This allows a flexible configuration on-site and enables the setup of test and development environments in Europe where only certain pieces of hardware are available.

In addition to the on-site database there are duplicates in four different locations in Europe. These replicated databases are also used for off-line quality checks and analysis.

The \hess{} DAQ uses a database abstraction layer, called \emph{simpletable}, to access the database. It was designed and implemented as part of the \hess{} DBTools \cite{dbtools} (in C/C++) and additionally implemented in Python for the Central DAQ. \emph{simpletable} facilitates the grouping of multiple records in so-called \emph{sets}. Such a set is for example a collection of calibration parameters for all pixels of a telescope. The library takes care of table locking and transactions, so that either a complete set is written/modified or nothing at all.

Another important aspect is that a  permanent history of configurations, calibration parameters etc. is provided. It is possible to store and access multiple versions of a configuration set. Under normal circumstances only new sets of data are added, even when these are just minor modifications of older sets still residing in the database. This also makes it easy to create temporary configurations, e.g.~deactivating a malfunctioning piece of hardware during the night and to roll back to an earlier version of the configuration when the error is resolved.

\subsection{Logging and error propagation}\label{software:errors}

A dedicated logging framework has been implemented in the DAQ. Log files for every DAQ process are created on a daily basis and timestamps with microsecond precision are written with each message to file. These logs are easily accessible using the central DAQ GUI and are stored for several years on-site and occasionally copied to Europe.

\begin{figure*}[!htb]
\includegraphics[width=\textwidth]{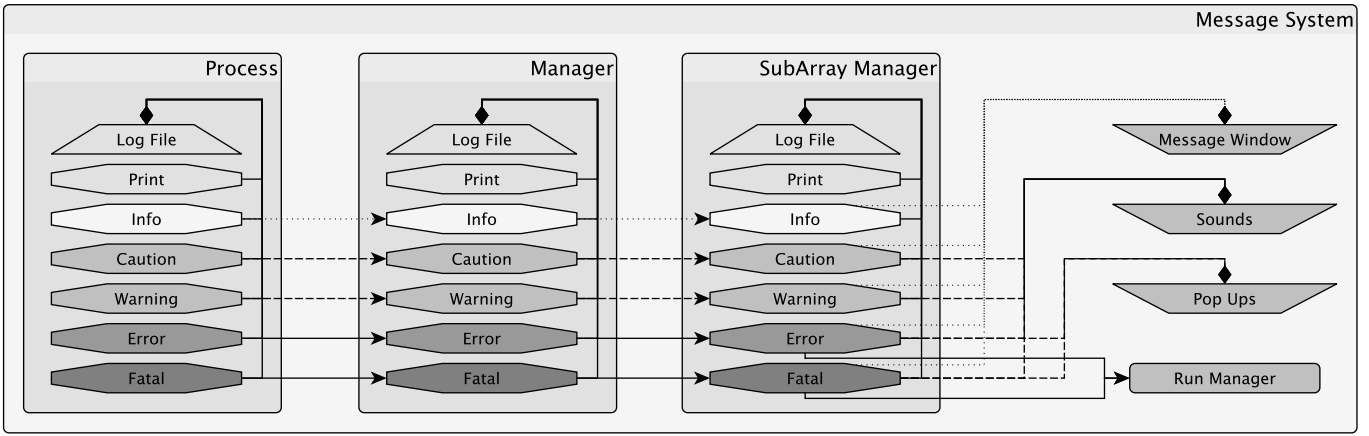}
\caption[]{There are six distinct message types used in the \hess{} DAQ system for inter-process communication and logging. A \emph{Print} message is piped into the daily log file of the corresponding process and is not distributed any further. An \emph{Info} message is send upwards in the hierarchy of the DAQ processes and also printed in the DAQ log message window which is displayed on one of the screens of the main DAQ control PC in the control room. \emph{Caution} and \emph{Warning} messages as well as \emph{Error} and \emph{Fatal} messages are also distributed upwards in the DAQ process hierarchy. Furthermore, all four of these message types also generate different sound notifications in the control room to get immediate shifter attention. \emph{Error} and \emph{Fatal} messages can also generate a pop-up window on the central screen of the main DAQ control PC. Moreover, the Run \emph{Manager} is notified if an \emph{Error} or \emph{Fatal} message was issued and no further runs are scheduled, i.e.~the \emph{Run Manager} is doing a transition to the \emph{Safe} state. Also the corresponding \emph{SubArray} is send to the \emph{Ready} state in case of an \emph{Error} message and the \emph{Safe} state in case of a \emph{Fatal} message as a safety precaution.}
\label{Messages}
\end{figure*}

There are six different message types available in the logging framework as shown in Figure~\ref{Messages}. Possible actions on message reception can range from print-outs to the log files, to pop-up messages with or without sound for the Shifters, to an automated security shutdown of the whole telescope array by the DAQ.

The log files are primarily used by the Shift Crew during error recovery to find the source of the underlying problem. Furthermore, the corresponding subsystem experts can use the log files at any time to search for problems. 

If a severe error is detected by a hardware component, it has to be able to prevent any damage to itself and inform its \emph{Manager} via its \emph{Controller} about the error state after immediate danger is averted. Once the DAQ is informed about the malfunction a fully automated procedure takes over which, depending on the severity of the error, can bring the whole array to a safe and consistent state. This automatic procedure is designed to prevent possible damage to any other hardware equipment but especially to take care of human safety, i.e.~to stop the telescope movement immediately, to shut down the Cherenkov Camera HV, etc. If an \emph{Error} does occur the corresponding \emph{SubArray} is brought to the \emph{Ready} state as quickly as possible using so called “immediate transitions”. They work exactly like normal transitions with the exception that all dependencies are ignored ensuring the arrival into a safe state as fast as possible. In case of a \emph{Fatal} error message the \emph{SubArray} is sent to the \emph{Safe} state and all hardware belonging to the run is shut down. Once the DAQ is finished with its automatic response, the Shift Crew has to take over and identify and solve the problem to be able to continue with normal operation.

\subsection{DAQ start \& stop}

Another highly automated procedure is the start and shut down of the DAQ. Starting the DAQ can be done in less than $\unit[1]{min}$ and stopping in less than $\unit[0.5]{min}$ on average. The complete DAQ system can be started or stopped using a single button in the central GUI or running a single command on the shell in a matter of seconds. This starts the \emph{Resource Handler}, which gets the list and configuration of all the processes that belong to the DAQ from the MySQL database and starts \emph{Host Handlers} on every machine that has been configured to be part of the DAQ in the database. These handlers are responsible for starting, monitoring and, if necessary, restarting all the processes that run on their machine.


\section{Array operation}\label{ArrayOperation}
The \hess{} Array takes data in periods of $\unit[28]{min}$, so called \emph{runs}, during astronomical darkness only. There are \emph{observation} runs which make up the bulk of the available dark time. Furthermore, dedicated \emph{calibration} runs\footnote{\emph{Calibration} runs are needed to determine the conversion of the recorded electrical signal into a number of Cherenkov photons. A more detailed description of the \hess{} calibration is given in \citep{hesscameracalib}.} are taken with the detector at regular intervals as well as other special purpose runs, e.g.~system tests at the beginning of the night. All the different run types are specified in a database and run types can be added, removed or modified without the need to change any code. This includes any combination of hardware that has to take part in a run of a given type. This is also true for the detailed configuration parameters (\emph{Run Parameters}) of the hardware used in a given run (which can be different for different run types).

Furthermore, the targets scheduled for observations during a given shift are contained in the same database and a dedicated tool, called the “AutoScheduler” \cite{autoscheduler} schedules all \emph{observation} runs for a given night. The AutoScheduler takes into account various predetermined conditions, e.g.~target priority, zenith angle, number of runs already taken on that target and available telescopes and uses an optimisation algorithm to prepare the schedule. This schedule is then written to the database and processed by the DAQ. The Shift Crew can adjust the schedule, e.g.~by adding \emph{calibration} runs manually, but is not allowed to change the observation schedule, unless there are exceptional circumstances, e.g. a ToO alert.

For further flexibility the DAQ can schedule runs for any combination of available telescopes. This includes multiple runs with different sets of telescopes, for example, an observation run using CT1, CT2 and CT4, a calibration run with CT3 and another observation run with CT5 on a different target. For that purpose, \emph{SubArrays} are used to manage the participating hardware in a given run. In the example above, three \emph{SubArrays} would have been used to take data with all five telescopes in three different runs at the same time.

The execution of the different scheduled runs is done by the \emph{Run Manager}. It constantly monitors the available resources (e.g.~telescopes and nodes) and is aware of the scheduled runs, their type and their requested hardware. It sequentially parses all scheduled runs and checks for free resources. As soon as all requirements are fulfilled (mainly there being enough free resources) the \emph{Run Manager} allocates a \emph{SubArray}, and sends the run configuration to the corresponding \emph{SubArray Manager} (see Section \ref{Manager}) which will then proceed on its own, i.e.~configure all the necessary hardware, start the data taking and, once the run is finished, un-configure the hardware again.

\subsection{User interface}
\begin{figure*}[!htb]
\includegraphics[width=\textwidth]{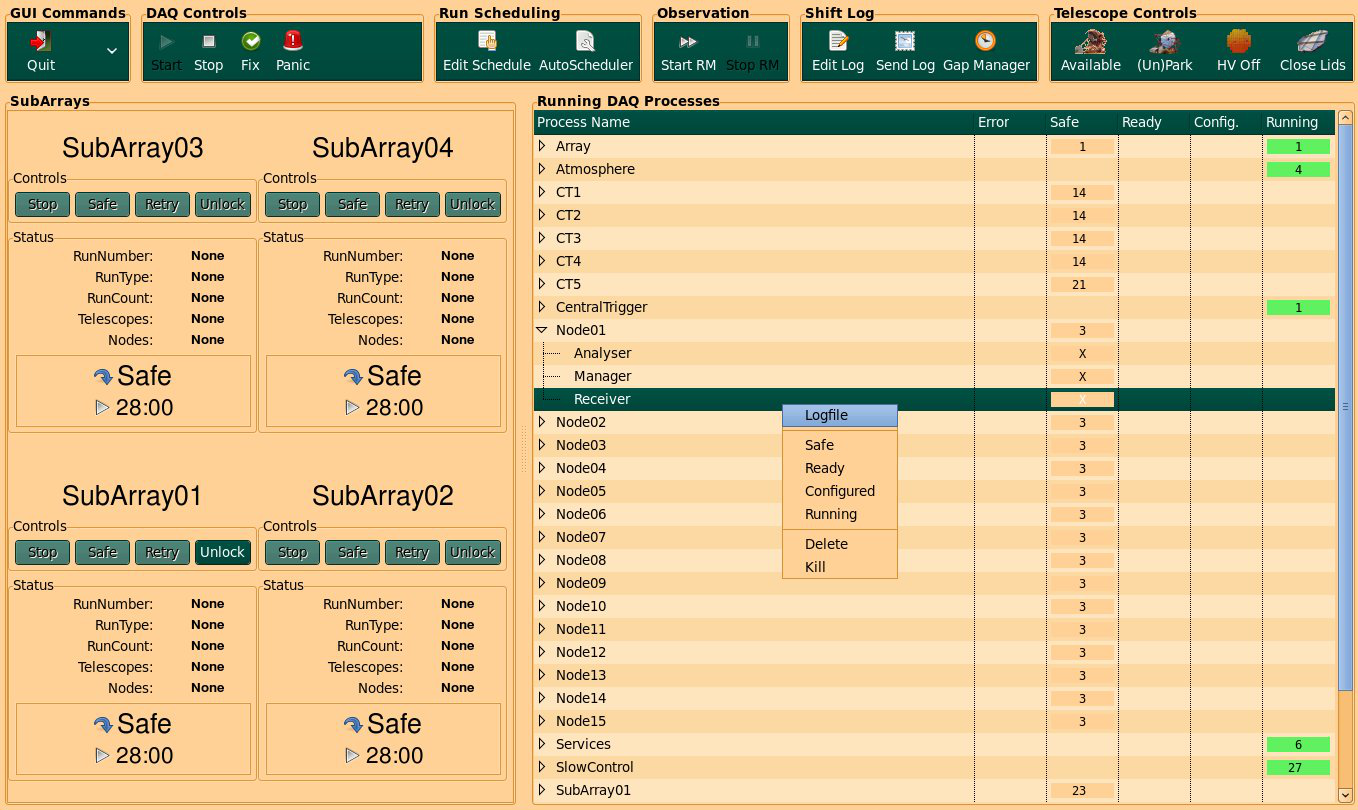}
\caption[]{Main part of the central DAQ GUI used by the Shifters to interact with the \hessII{} DAQ. It is separated in three distinct parts. The upper row consists of several groups of buttons that are needed during normal operation. The lower left part of the GUI is dedicated to giving detailed information about and control over the available \emph{SubArrays}. This includes information about run duration, observation target, used telescopes, etc. The main part of the GUI is used to show all running DAQ \emph{Controllers} to the Shifters. They are grouped in contexts and can be expanded to show all processes within this context. The amount of processes in a context as well as their current state are also shown. Detailed control over every single process (even multiple ones if more than one are selected) is possible using the right-click menu. In case of errors, the \emph{Controllers} at fault are marked with an error flag (not shown).}
\label{DAQGUI}
\end{figure*}

All of the interaction between the Shift Crew---as well as other subsystem experts---and the DAQ is done in the central control room on-site. Several dedicated display machines are used to show monitoring information to the Shifters as well as to give fast feedback about the current status of the array. This includes information about the weather conditions outside (temperature, air pressure, wind speed and humidity), the camera monitoring (temperature, high voltage and currents), telescope pointing and motion as well as real time calibration and analysis data (e.g.~shower images in the cameras and sky maps containing the source direction of all reconstructed gamma-like events). A central graphical user interface (GUI) is used for interaction with the DAQ; a screenshot of the main part of the GUI is shown in Figure \ref{DAQGUI}. It serves as a single point of contact between Shifters and the DAQ where all essential settings and configuration options can be changed. This allows Shifters to take over manual control in case of error recovery and special operations in an easy way. The central GUI, and most other GUIs in the \hess{} DAQ, are implemented in the Python programming language using PyGTK \citep{pygtk}. Furthermore, all GUI processes implement the state \emph{Controller} interface and are managed like all other processes in the DAQ, e.g.~they receive \emph{Run Parameters} and perform transitions like \emph{Starting} and \emph{Stopping}.

\begin{figure*}[!htb]
\includegraphics[width=\textwidth]{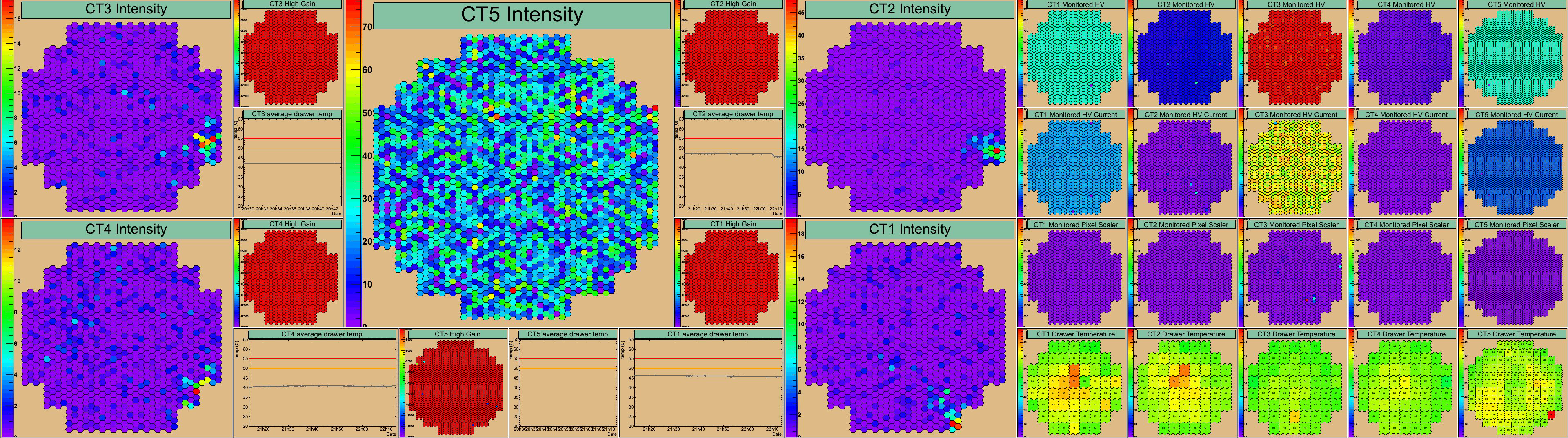}
\caption[]{Example selection of slow-control displays for immediate feedback to the Shift Crew. The five largest camera displays show pixel intensities in photo-electron volts for all telescopes. Right next to the intensity plots the High Gain ADC count camera displays as well as the average camera drawer temperature over time histograms can be seen. On the right side of the screenshot the different rows of camera displays from the top to the bottom correspond to current pixel HVs, pixel currents, pixel scaler values and drawer temperatures (the pixel scaler value being the number of pixel triggers in a certain time period for a given pixel).}
\label{Displays}
\end{figure*}

Building upon the common raw data format based on ROOT, a generic data \emph{Displayer} has been developed (using C++). It uses the object introspection capabilities of the ROOT data analysis framework to gain access to any data member of any \hess{} data storage format. Therefore it is able to plot any data that are recorded in the \hess{} DAQ and can be configured solely using the MySQL database. Different specialisations of this \emph{Displayer} can plot time lines, bar charts, wind roses, camera images, etc. An example of available displays that are shown in the control room can be seen in Figure~\ref{Displays}.

Some basic data quality checks can be performed with the \emph{Displayer} as well, e.g.~range checks with warning sounds and pop-up messages. Normally, displays are updated at a rate of a few Hz providing fast feedback to the Shifters.

For complex hardware components, e.g.~the Cherenkov cameras, expert GUI modes are also available. Those allow detailed control of the various hardware components and can be used by subsystem experts as well as experienced Shifters on-site for development, error handling and debugging.

\subsection{Real-time pipeline}\label{RealTimePipeline}

A dedicated collection of Displayers is also used to show the plots of the \enquote{real-time pipeline} to the Shifters. It is a full analysis, based on the HAP TMVA analysis \citep{HapTmva}, of the data being taken running in real time. The only limitation is that a default camera calibration has to be used, e.g.~the High Gain to Low Gain ratio, flatfield coefficients, single photo electron values as well as muon coefficients with the exception of the calculation of the pedestal position of the two gain channels of the photomultiplier~\citep{hesscameracalib,Funk,Balzer}. For background determination either the Ring Background (the default; using several speed-ups concerning coordinate transformations), the Reflected Region or the Template Background method can be used~\citep{HessBackground}. Run results are stored on disk and used to perform a real-time analysis with data from consecutive runs on the same target. The output of the real-time pipeline are calibrated camera images with intensities in photo electrons as well as significance maps of the region of the sky that is currently being observed and $\theta^2$ plots around the target source position. If a significant detection of a source is made the Shift Crew is alerted using pop-up messages and advised to call for expert input to allow swift follow-up observations. To cope with the high-data rates the real-time pipeline is split into several different processes. Each \emph{CameraReader} has a corresponding \emph{Analyser} process which subscribes to the data stream that is processed by the \emph{CameraReader}. The \emph{Analyser} processes all events that are generated by the \emph{CameraReader}. This includes event calibration and event reconstruction as well as gamma-hadron separation. The processed data is collected by \emph{AnalysisServer} processes; one for each \emph{SubArray} in use. The input maps for the significance maps are filled at the \emph{Analyser} and the final significance maps are created at the \emph{AnalysisServer}. The latter is a time consuming processes (several minutes) and therefore happens in parallel to the input maps being received.

\subsection{Monitoring and shift logs}\label{tt_tools}

To calculate the data taking efficiency of the \hess{} telescope array the so called Transition Time Tools are used. It is a Python framework that analyses the gaps between the various runs and the timestamps, that were written to a database by the different DAQ \emph{Controllers}. The DAQ monitors whether there are any gaps in data taking during dark time. If a gap is detected the Shifters are asked to give a reason for each gap between data taking runs. The answers the Shifters provide are stored in the same database. In combination, the information about the transitions, the gaps between runs and their reasons can be used to calculate the data taking efficiency of the \hess{} Array as well as the percentage of lost observation time of the array due to DAQ problems. Furthermore, it is possible to benchmark different processes with microsecond resolution and identify bottlenecks in the data-taking procedure. A further source of information is the detailed shift log that is sent to a mailing list and that is stored on disk after every night of a shift to keep the collaboration up to date about the ongoing activities on-site.

\subsection{Error handling}\label{ErrorHandling}

In case of a hardware failure, the DAQ automatically performs a safety shutdown, as described in section \ref{software:errors}. Apart from this automatic procedure, manual overrides are available for every automatic configuration and automatic DAQ action. If a hardware device is not available or malfunctioning, its corresponding \emph{Controller} can be replaced with a \emph{noopController} which just implements the common interface of the state machine while not doing anything else. This procedure is wrapped in a Python script with the name of the \emph{Controller} that should be replaced as a single command line parameter allowing the Shift Crew to easily continue observations without the failed hardware.

Along with the manual overrides, extensive documentation of all hardware components and the DAQ are available on-site. While the hardware manuals are mostly present in analogue form, the DAQ manual and the DAQ troubleshooting guide are located within a MediaWiki \citep{wiki:mediawiki} on-site. Shifters are encouraged to contribute to the Wiki while they are on Shift with an emphasis on the Shifter’s Notes, a summary of the current shift for the next Shift Crew.

At the beginning of each Shift the current Shift Crew is introduced to the handling of the array and to the emergency procedures by a local Shift Expert who spends the first ten days with the Shifters. After that time the Shifters are on their own. Should they be unable to solve a problem, subsystem experts on call are available to help resolve the problem.

\subsection{Remote control}

The \hess{} site is located in a remote region with very limited internet connectivity. To make the remote maintenance easier and to minimise the number of maintenance trips from Europe, several remote maintenance tools are used. For instance, IPMI-cards \citep{ipmi} are installed on every machine in the DAQ cluster. This allows the machines to be power-cycled remotely as well as giving access to the BIOS and other configuration menus during the boot-up of the machines. Furthermore, VNC \citep{wiki:vnc} servers are used to forward the graphical displays once the operating system has started. They can be started on every machine of the DAQ cluster and are running constantly on the machines in the control room. The VNC connections have proven to be an invaluable tool when it comes to remote assistance in case of problems during data taking as well as for remote maintenance of the cluster and related machines. The access to these VNC servers is restricted to DAQ experts to prevent tampering with the system by third parties.

The remote access to the network is possible with an OpenVPN \citep{openvpn} server running on the gateway machine of the DAQ cluster. With this it is possible to access all of the different networks on-site without compromising security or the separation between the data network and users network. To ensure a stable production environment, the Shift Crew as well as other members of the collaboration cannot change the software used for data taking. Only DAQ experts can make software changes and can, therefore, guarantee a properly working DAQ system. The Shifters are given restricted access to the cluster to minimise the possibility of human error.

\subsection{Reaction to ToO alerts}
The DAQ system can receive target of opportunity alerts from other experiments via the Gamma Ray Burst (GRB) Coordinates Network (GCN). A dedicated process, the \emph{GCNAlerter}, has been developed by the \hess{} Collaboration. It listens for messages from the GCN network, checks whether the coordinates are visible and takes further action. For \hess{} Phase~I the \emph{GCNAlerter} informed the Shift Crew and prepared a script to alter the observation schedule, which had to be confirmed and executed by the human operators. The results of the Gamma Ray Burst observations with \hessI{} are described in \citep{hessgrb}. For the start of \hessphaseII{} a revised target of opportunity alert scheme has been developed by the \hess{} Collaboration \cite{newhessgrb} and is currently being implemented into the DAQ. If the \emph{GCNAlerter} decides that an alarm justifies a prompt observation, the DAQ will react fully automatically and start data taking on the new target without the need for human intervention. For safety reasons the Shift Crew is not allowed to enter the array if the ToO alert system is active. 

For these prompt observations the time span between receiving the alert and the beginning of the observation has to be minimised; transient events occur on time scales of a few seconds and several minutes. The bulk of the transition time between two runs is due to the slewing time of the telescopes to the new target. In normal operation, some additional time is used to switch off the high voltage while the telescopes are moving to prevent damage to the photo multipliers of the Cherenkov cameras due to bright stars in the field of view. 

In case of a GCN alert, the telescopes immediately start moving to the new target while the ongoing runs are stopped and the reconfiguration of all \emph{Controllers} is done. Moreover, the high voltage of the camera is not turned off and the cameras are just sent to an internal paused state where they stop taking data but remain fully configured. The effect of these actions is that the transition time is just the slewing time of the telescopes which cannot be sped up any more (see \cite{CT5tracking}) and a short overhead for the unpausing of the camera and the reactivation of the camera pixels.

Moreover, dependencies which are optional processes are not waited upon so that they do not increase the duration of the transition. On top of that, only CT5, due to its higher movement speed, is required for the start of the data taking, CT1 to CT4 are configured to be optional processes and will join data taking once they are on target. The Cherenkov Camera Trigger and Central Trigger configuration are the same as during normal observations (CT5 monoscopic trigger are allowed all the time). The drive system of CT5 also has the option to use reverse tracking of ToO targets, i.e.~driving beyond zenith, and the fine positioning of the telescope is done during data taking because the errors on the target position of a ToO alert are large.

\section{Software management}

\subsection{Test-DAQ cluster \& DAQ simulation}

To test the DAQ software without real hardware, a full DAQ simulation can be set up, i.e.~raw camera data are sent from a camera emulation process to the node receivers and a Central Trigger emulation process sends trigger blocks to the corresponding receivers as well. This can be used to test receivers, the node switching as well as the real time analysis with real data as input. However, to process a full run of real data in an acceptable time frame, a computer cluster similar to the one in Namibia has to be used. The \hess{} Test-DAQ-Cluster, a scaled down version of the DAQ cluster on-site (five nodes instead of ten, two storage servers instead of five, one switch instead of four), provides enough computing power for that purpose. Moreover, the same operating system and software is running on the Test-DAQ, making it an ideal test bench for software development. Its location in Europe also allows easy access in contrast to the cluster on-site.

\subsection{Development tools}

During the development of the \hessII{} DAQ software the Make-based build system \citep{make} of the software was replaced by one using SCons \citep{scons}. Apart from a code cleaning during this transition, it is now possible to build the software using multiple jobs in parallel. The legacy build system was not able to build the software correctly using multiple jobs. It was simpler to re-implement the build system, instead of modifying the Make-based one\footnote{The SCons built-in dependency management is more sophisticated and can be extended quite easily compared to Make's built-in dependency rules.}. Another benefit of SCons is the use of Python for its configuration script files which allows a quick start for beginners and facilitates maintenance of the build system. To further aid in development a Bugzilla Bug tracker \citep{bugzilla} is used by the software developers of the \hess{} Collaboration.

For development purposes and benchmarking so-called \emph{Dummy Controllers} are available in C++ and Python. They provide the basic interface of the state machine and are used to mock missing hardware \emph{Controllers} in a testing environment. The Dummy Controllers are also used to test the dependency and time-out handling within the DAQ as well as to test the automatic shutdown and mechanisms in case of errors. They mock long-running transitions or can throw exceptions during specified transitions, depending on the options that are passed as command line arguments.

To aid other subsystem experts in the development of their own DAQ \emph{Controllers} for their hardware a dedicated virtual machine, the DAQ-VM, is available. Most of the time the subsystem hardware cannot be shipped to the Test-DAQ-Cluster and a software test environment has to be set up at the location of the hardware (for instance properly configured operating system, database and CORBA omniName server). The DAQ-VM was created using VMware Fusion 3 \citep{vmfusion}, the hardware requirements are a single-core $\unit[2]{GHz}$ processor, $\unit[1]{GB}$ memory and $\unit[20]{GB}$ disk space. Taking the hardware requirements into account, the DAQ-VM can be run on almost all currently available laptops. This allows non-DAQ-Experts to test new DAQ \emph{Controllers} with their corresponding hardware under conditions which are as close as possible to those on-site without detailed knowledge about how to set up such a test environment.

Another helpful tool for non-DAQ-Experts is the detailed documentation available in an internal Wiki of the \hess{} Collaboration. Together with basic example \emph{Controllers} and the extensive how-to guides in the Wiki non-DAQ-Experts can quickly start to write and test new DAQ \emph{Controllers}.


\section{Performance}
The \hess{} DAQ system has been in use since the commissioning of the first \hess{} telescope in 2003. Since then the DAQ system has evolved continuously to its current state as described in the previous sections. Over the 10 year period of operation corruption of data due to central DAQ malfunctions was extremely rare and is negligible. In preparation for \hessphaseII{} the central DAQ software has been overhauled, including being ported from $\unit[32]{bit}$ to $\unit[64]{bit}$, to make full use of the architecture of recent server machines. The performance of the central DAQ software will be presented in the following, focusing on the system that was prepared for \hessphaseII{}.

\subsection{Stability}
The \hess{} DAQ has been in operation for almost ten years and performed well throughout this time. In spite of frequently failing hard disks and the harsh environment for sensitive electronic equipment, the amount of data lost over time is negligible. The redundancy of the available hardware and software has played a critical role in this achievement, i.e.~broken hard disks are replaced immediately by the RAID setup and other broken components can be replaced with spare parts from a common pool. At the same time it is possible to redistribute processes to other machines, because they have identical hard- and software environments. This is especially true for all services needed by the central DAQ and all DAQ \emph{Controllers}, which can be started on any machine of the cluster (storage-server or farm node). This multi-level redundancy design both minimises the probability of losing data as well as the recovery time after a computer hardware failure.

The central DAQ overhead contributing to the transition time between two consecutive observation runs is negligible, i.e.~the time needed by the hardware to be ready for the next run is several orders of magnitude bigger than the central DAQ overhead of the corresponding controllers. Moreover, the central DAQ does not increase the system deadtime. In the time period from 01.01.2009 to 31.12.2012 the \hess{} Array was not operational due to central DAQ problems for $\unit[0.8]{\%}$ of the available dark time. This demonstrates that problems with the central DAQ hardware and software during data taking can be quickly solved.


During data taking the \hessII{} Array produces a total data rate of $\unitfrac[46]{MB}{s}$ on average. This includes $\unit[1]{\%}$ of slow control data and of log files which are stored every night for debugging purposes. The average CPU load of the cluster during data taking does not exceed $\unit[10]{\%}$ (of five $\unit[2.5]{GHz}$ quad cores and ten $\unit[3]{GHz}$ octo cores), while the average memory usage is below $\unit[30]{\%}$ (of a total of $\unit[200]{GB}$). In a month worth of data-taking on average $\unit[11]{TB}$ of data are written to disk, which allows several months of observation data to be stored on disk in Namibia. Overall the performance of the central DAQ cluster is more than sufficient to handle the data rates of the \hessII{} Array, leaving enough room for computationally intensive tasks, like e.g.~preliminary analysis and elaborate data quality checks.

In addition to the slow-control data, real-time data quality checks and real-time analysis results provide fast feedback about the current status of the array and about the scientific quality of the running observation. Shifters have easy access to these data using the different screens in the control room, allowing immediate intervention in case of problems.

The operation of the \hess{} Array by non-expert personnel is possible due to the detailed manuals and documentation about every hardware subsystem and the high level of automation within the DAQ. This automation and the open-source remote administration and monitoring tools greatly reduce the maintenance cost of the \hess{} central DAQ software, which currently requires approximately a $\unit[0.5]{FTE}$ position in Europe.

\subsection{Flexibility}
New hardware components can be added easily to the \hess{} Array. Only the software for the \emph{Controller} responsible for the new hardware has to be written and added to the DAQ. The rest of the system, i.e.~data transport, storage and visualisation does not need to be modified. Furthermore, the inter-dependence of the new \emph{Controller} with already existing ones have to be added to the database. 

The central DAQ software used for \phaseII{} of the \hess{} telescope array evolved from the central DAQ software designed for Phase I of the experiment. One of the main differences between the new central DAQ software and the earlier implementation is compatibility with $\unit[64]{bit}$ architecture of recent CPUs, which required many minor patches. However, the principal design ideas of the \hess{} central DAQ software remained the same. This includes the configuration of the DAQ itself and the controlled hardware from a central database, as well as the common data format for all monitoring and scientific data. For the latter only additional information had to be  included, i.e.~the-pixel wise timing information that became available with the newer electronics \citep{CT5camera} of the CT5 camera. The fact that no redesign of the central DAQ software was necessary is due to the uniform interfaces for hard- and software: Ethernet standard for communication to all hardware components and the common interface of the software \emph{Controllers} for all devices.

Another example of the capability of the \hess{} DAQ to quickly adapt to new situations, is the fact that during the whole commissioning of the fifth telescope the live system was used to test and develop \emph{Controllers} relevant for the new telescope while the array was taking data using the other four. This includes parallel and joined observations of the old and new telescope. Due to this ability to take data with different \emph{SubArrays} the CT5 commissioning could be done mostly in parallel to data taking with CT1-4 which had only minor downtime. 

The commissioning for \hessphaseII{} of the central DAQ went smoothly. This was possible due to the extensive documentation and guides about DAQ \emph{Controller} development, and the DAQ Virtual Machine (VM) which allowed non-DAQ-Experts to easily develop and test the \emph{Controllers} for their own custom hardware. For example, the development of the Tracking Controller for CT5, the \emph{Controller} for the calibration device of the new CT5 camera as well as the adjustment of the Central Trigger Controller has benefited. Moreover, the DAQ VM was used to test the central DAQ software with the camera test-benches (this includes old and new camera hardware, i.e.~test-benches mimicking CT1-4 as well as CT5) during the initial development phase of the DAQ for \hessphaseII{}.

The main platform for the development of the \hessphaseII{} central DAQ was the TestDAQ, including a full simulation of the data-taking process. In a simulated observation run, real raw data from previous runs taken with the \hess{} DAQ are fed into the system, which allows the central DAQ to be tested under near real conditions, the main difference being that there is no actual hardware connected to the TestDAQ cluster and \emph{Dummy Controllers} take the place of real ones. These \emph{Dummy Controllers} can also be used to fake slow hardware and to test the error handling of the DAQ. The simulation runs can be used to test the real-time analysis, to test data quality checks and to do benchmarks with the hardware of the cluster as well as the DAQ software running on it to identify bottlenecks. Moreover, using the TestDAQ to test firmware and driver updates, as well as other hardware or software modification, before applying them to the live system helps to avoid complications on-site and contributes to the stability of the DAQ.


\section{Conclusion}
The \hess{} central DAQ has been in operation for almost 10 years without any major problems since the inauguration of the first telescope in 2003. The central DAQ did only contribute to a loss of $\unit[0.8]{\%}$ of the available dark time since 2009 proving its ability to quickly deal with hardware or software problems during data taking. Moreover, the amount of data that have been lost is negligible.

At this point we would like to share the most important lessons learned which could be interesting for other experiments like e.g.~the upcoming CTA Project.

The ROOT-based data format proved to be useful as a common format for both on-line and off-line analysis software. However, the tight dependency on an external framework lead to various issues. The software had to be adapted to the three major releases of the ROOT-framework (v3, v4, v5) during the last years. This caused additional work in terms of software development (ensuring back-wards compatibility with existing raw data files and classes) and also in terms of system administration (maintaining different ROOT versions on different operating systems).

The distributed development of the central DAQ software and hardware \emph{Controllers} in the \hess{} Collaboration proved to be difficult due to the many different environments at  the different member institutes and laboratories. To avoid some of the problems that usually arise during updates on site (i.e.~introducing new functionality for \emph{Controllers}, installing updates for the operating system or cluster hardware drivers) it proved very usefully to integrate the software and test the installation on the TestDAQ and the DAQ Virtual Machine, which helped to resolve many problems beforehand.

While technically the central DAQ software is organized in separate modules, there are a lot of historic interdependencies between the different modules and classes. As a result, it is very difficult to effectively test the different components of the DAQ software without integrating the full software distribution (both DAQ software, and off-line analysis packages). The DAQ Virtual Machine helped to test \emph{Controllers} in a \enquote{standardized} environment. It provides a pre-installed reduced collection of \hess{} DAQ software modules and their dependencies for easy development and testing.  Thus it alleviated some of the mentioned problems. For future projects each software module should provide automated tests that can be run without the need to integrate other modules. This helps to ensure a modules functionality and also reduces the coupling between modules, making maintenance a lot easier.

The abstraction of internal hardware states to the simplified linear state machine proved to be very useful in terms of controlling the whole telescope array or various subsets. This is also true for the design decision to delegate the safety of each piece of hardware to its firmware, and only having one \emph{Error} flag in the \emph{Controller}, which tells the central DAQ whether a piece of hardware is working or not. With these simplifications it is easier to keep the whole DAQ in a consistent state and to use automatic procedures like the controlled emergency stopping of a run in case of hardware failures.

The virtualization of the legacy boot-servers (described in \autoref{storage}) for the Cherenkov cameras and Central Trigger devices has proven to be very useful. It allowed to replace the outdated special purpose machines with new off-the-shelf hardware while keeping the specialized operating system with all its customizations without any additional reconfiguration or development. For future projects it is a good idea to decouple the configuration of the DAQ system from the physical set-up of the cluster computing hardware. This could be achieved by creating virtual machines for all critical tasks (e.g. database servers, boot servers, computing nodes, etc.) and distribute them dynamically on the available physical machines. This would also reduce the time to reconfigure the system after a physical server or computing node failure during data-taking.

Another lesson learned concerns the graphical user interfaces in the control room. The control and monitoring interfaces are very valuable to the Shifters and enable them to take charge of observations after a few nights of training. On the technical side it would have been better to decouple the control functionality from the actual display code. Quite some issues had to be resolved involving multi-threaded interprocess communication mixed with multi-threaded graphics display code. Our suggestion for future projects is to provide user interfaces using standard web interfaces. The main advantage is that the system becomes almost independent of the actual display machines. The web interfaces can be displayed on any current or future operating system or device as long it provides a web-browser. Moreover, it would simplify remote operation and monitoring by providing remote access to the web interfaces allowing the same interactions with the system that would be possible on-site.

Due to the high flexibility and scalability of the design of the central DAQ, the inevitable changes to the central DAQ necessary due to upgrades of the array could all be resolved in an evolutionary fashion in contrast to a costly redesign. This is especially true in the light of the commissioning of the fifth telescope of \hessphaseII{}.

\begin{small}
\section*{Acknowledgements}
\noindent
The authors would like to acknowledge the support of their host institutions. 
We want to thank the whole H.E.S.S. Collaboration for their support.

The support of the Namibian authorities and of the University of Namibia
in facilitating the construction and operation of \hess{} is gratefully
acknowledged, as is the support by the German Ministry for Education and
Research (BMBF), the Max Planck Society, the German Research Foundation (DFG), 
the French Ministry for Research, the CNRS-IN2P3 and the Astroparticle Interdisciplinary Programme of the
CNRS, the U.K. Science and Technology Facilities Council (STFC),
the IPNP of the Charles University, the Czech Science Foundation, the Polish 
Ministry of Science and  Higher Education, the South African Department of
Science and Technology and National Research Foundation, and by the
University of Namibia. We appreciate the excellent work of the technical
support staff in Berlin, Durham, Hamburg, Heidelberg, Palaiseau, Paris,
Saclay, and in Namibia in the construction and operation of the
equipment.
\end{small}

\bibliographystyle{model1-num-names}
\bibliography{main}
\end{document}